\documentclass[prb,twocolumn,showpacs,superscriptaddress,aps,amsfonts]{revtex4-1}
\usepackage{graphicx,graphics,color,epsfig}
\usepackage{bm}
\usepackage{amsmath}
\usepackage{amssymb}
\usepackage{epstopdf}
\usepackage[mathscr]{eucal} 

\begin{document}
\bibliographystyle{prsty}
\title{Stripes in high temperature superconductors dilate under\\ the influence of an external electric field }

\author{Daniel SPRINGER}
\email{spri0001@ntu.edu.sg}
\author{Siew Ann CHEONG}
\email{cheongsa@ntu.edu.sg}
\affiliation{Division of Physics and Applied Physics, School of Physical and Mathematical Sciences, Nanyang Technological University, 21 Nanyang Link, Singapore 637371, Republic of Singapore}

\date{\today}

\begin{abstract}
In this work we study the influence of an electric field on stripes and compare our results with experimental findings from scanning tunneling microscope measurements. By introducing a negative-bias electric field into a time-dependent Ginzburg-Landau equation with stable stripe solutions, we show that hole(electron)-like stripes widen (narrow) compared to the field free case. When a magnetic field is introduced instead, stripe formation is found to be suppressed. \\
PACS: 74.20.De, 89.75.Kd, 74.20.-z 

\end{abstract}

\maketitle

Theoretically stripes in high-temperature superconductors (HTSC) have been predicted by Emery and Kivelson in Refs. \onlinecite{Kivelson1990, FirstStripes, KivelsonStripes}. Based on their theoretical study on electrons hopping in an antiferromagnetic background, Emery and Kivelson found that it is energetically favorable for electrons to organize themselves spontaneously into stripes. Later White and Scalapino along with Lorenzana and Seibold gave further theoretical support to the existence of HTSC stripes in their density-matrix renormalization group (DMRG) \cite{Scalapino1998} and mean-field \cite{Lorenzana1998} studies.

Early experimental evidence for the existence of stripes came from neutron scattering studies in La$_{1.6-x}$Nd$_{0.4}$Sr$_x$CuO$_4$ by Tranquada et al. \cite{Tranqada1995}.
In their later STM measurements of Bi$_2$Sr$_2$CaCu$_2$O$_{8+\delta}$ \cite{Hoffman}, Hoffman et al. found density of states (DOS) modulations that widen along the $(\pi, 0)$ direction and narrow along the $(\pi, \pi)$ direction, for increasingly negative bias voltage.
Instead of identifying these DOS modulations as stripes, Hoffman et al. concluded that these are quasiparticle interference patterns, based on comparisons of their STM results with ARPES data.
However, in a review on fluctuating stripes in high-temperature superconductors\cite{KivelsonModRev}, Kivelson et al. argue that the detection of quasiparticles does not rule out the existence of stripes.

In general, STM experimentalists assume implicitly --- as Hoffman et al. did --- that the electric fields generated by their STM tip are too weak to influence the surface being probed.  However, since stripes are ultimately aggregations of charge carriers \cite{KivEmTra}, it is much more reasonable and intuitive to expect the opposite.  In this work, we explore how a charged stripe would respond to external electric fields, and compare our findings with the observations in Ref. \onlinecite{Hoffman}.

To do so, we can start from a lattice model, for example the Hubbard model or the $t$-$J$ model with stable stripes and then introduce an electric field. However, in general it is not possible to solve such models analytically. 
Approximate solutions can be obtained but their accuracy cannot be guaranteed, whereas exact numerical solutions can only be obtained for small systems at a large computational cost.

An alternative approach relies on the descripion of the macroscopic state of a system by an order parameter.
Such order parameters characterize the thermodynamic behaviors of the system of interest, and obey Ginzburg-Landau equations (GLE).
GLEs without stripe solutions have been used to describe superconducting systems \cite{TangWang, Cyrot, RoLi}.
Of the various equations with stable stripe solutions, the complex Ginzburg-Landau equation (CGLE) 
\begin{equation}
\partial_t \Psi = \Psi + (1+i\alpha)\nabla^2 \Psi - (1+i\beta)|\Psi|^2 \Psi.
\label{Equation1}
\end{equation}
introduced by Newell and Whitehead \cite{NewWhit} is one of the most well-studied nonlinear equations in the pattern formation community.
Here $\alpha$ and $\beta$ are phenomenological parameters.
It has been used to describe phenomena from nonlinear waves to second-order phase transitions, from superconductivity, superfluidity, and Bose-Einstein condensation to liquid crystals and strings in field theory (see review by Aranson and Kramer \cite{WorldOfCGLE}) . 
In general GLEs have very similar structures and therefore, 
even though the CGLE was not derived from a microscopic model of superconductivity, we expect that qualitative conclusions derived from the CGLE also apply to other GLEs.

External electric and magnetic fields are introduced into Eq. \ref{Equation1} through minimal coupling
which then becomes 
\begin{equation}
\begin{aligned}
\left(\partial_t + \frac{iq}{\hbar}\phi\right)\Psi &=
\Psi + (1+i\alpha)\left(\nabla - \frac{iq}{\hbar}\vec{A}\right)^2 \Psi - {} \\
&\quad\ (1+i\beta)|\Psi|^2 \Psi,
\end{aligned}
\label{Equation2}
\end{equation} 
where $\phi$ is the scalar and $\vec{A}$ is the vector potential. 
We will use the effects of magnetic fields on stripes to demonstrate that it is plausible to associate $\Psi$ with the superconducting order parameter. 

We solve Eq. \ref{Equation2} numerically. For this nonlinear partial differential equation, spectral methods \cite{Trefethen, Boyd} are the most stable and accurate, where they are applicable. 
In these methods, we first Fourier transform the partial differential equation, to get a coupled system of non-linear ordinary differential equations of the form $ \dot{u} = cu + F(u)$, where $c$ is a constant and $F(u)$ a non-linear term.
These equations are then integrated using the exponential time differencing scheme \cite{Cox}
\begin{equation}
u(t_{n+1}) = u(t_n)\, e^{ch} + e^{ch} \int_0^h e^{-c\tau} F(u(t_n+\tau))\, d\tau,
\end{equation} 
where we approximate the non-linear term as
\begin{equation}
F = F_n + \tau (F_n - F_{n-1})/h + O(h^2). 
\end{equation}
This pseudo-spectral technique \cite{Winterbottom} works well if we are interested in field-free solutions or when only an electric field is applied. If a vector potential is introduced the differential operator does not reduce to a multiplication in Fourier space and thus spectral techniques are not convenient anymore.
To integrate the CGLE in the presence of magnetic fields we use a finite-difference approach \cite{FD} with Adam-Bashforth timestepping of third order instead\cite{footnote}. In our finite difference approach we take care of the vector potential by applying the gauge-invariant $\psi U$ method \cite{BergerRubinstein}. By introducing phase factors of the form
\begin{equation}
\mathscr{U}^x = \exp\left\{-i \int_{x_0}^x \textbf{A}_x(\epsilon,y)\, d\epsilon \right\}
\end{equation}
\begin{equation}
\mathscr{U}^y = \exp\left\{-i \int_{y_0}^y \textbf{A}_y(x,\mu)\, d\mu \right\} 
\end{equation} 
we find the simple equality
\begin{equation}
(\nabla - i\textbf{A})^2 =  \bar{\mathscr{U}}^x \partial_{xx}(\mathscr{U}^x \psi) + \bar{\mathscr{U}}^y \partial_{yy}(\mathscr{U}^y \psi)
\end{equation}
with $\bar{\mathscr{U}}$ as the conjugate of $\mathscr{U}$.
Because of the discrete lattice spacing, it is necessary to define link variables
\begin{equation}
\textit{U}^x_{i,j} = \bar{\mathscr{U}}^x_{i,j} \mathscr{U}^x_{i+1,j}, \quad \textit{U}^y_{i,j} = \bar{\mathscr{U}}^y_{i,j} \mathscr{U}^y_{i,j+1}.
\end{equation}
between two neighbouring points.
The Laplacian in our gauge-invariant finite difference approximation then reads 
\begin{equation}
\begin{aligned}
&\quad\ \left. (\nabla - i\textbf{A})^2 \psi \right|_{(x_i,y_i)} \\
&= \frac{\textit{U}^x_{i,j} \psi_{i+1,j} - 2\psi_{i,j} + \bar{\textit{U}^x}_{i-1,j} \psi_{i-1,j}} {a_x^2} + {}\\    
&\quad\ \frac{\textit{U}^y_{i,j} \psi_{i,j+1} - 2\psi_{i,j} + \bar{\textit{U}^y}_{i,j-1} \psi_{i,j-1}}{a_y^2}.
\end{aligned}
\end{equation}

\begin{figure}[t!]
\centering
\includegraphics[width=0.35\textwidth]{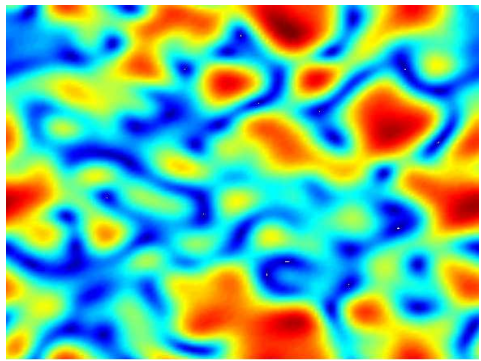}
\caption{When we start from random initial conditions the microscopic superconducting pockets act as nucleation seeds. These islands of homogenous phase grow up to a critical size before they break up into stripes.  The eventual width of stripes formed depends on the maximum size the islands reach before collapsing.}
\label{UniIslands}
\end{figure}

To understand how stripes form in the absence of external fields we start from random initial conditions for $\Psi$.
This describes a superconducting material slightly above the critical temperature $T_c$.
In this state microscopic superconducting pockets with finite lifetimes have formed everywhere.
As the simulation progresses we see that these microscopic pockets act as nucleation seeds, from which small islands of the homogeneous superconducting phase grow (see Fig. \ref{UniIslands}).
Upon reaching a critical size, these islands break up abruptly and evolve into the stripe phase.

The stability of stripes in relation to a phase-separated state in which a charge-density wave and a dilute liquid state coexist has been analysed in Refs. \onlinecite{HenleyZhang, HenleyZhang2} for spinless fermions in a Hubbard-like model. In this work the stripe phase was found to be marginally more stable than the phase-separated state at higher hole-doping.
However, to the best of our knowledge there has been no systematic study on the relative stability of the two phases in small droplets, and their associated finite size effects.
In addition to stripes the CGLE also allows the formation of hexagonal patterns. A necessary but insufficient condition for stable hexagon solutions is for the inversion symmetry $\Psi = \rightarrow -\Psi$ to be broken \cite{TsiAra, TsiAra2}. 
However, we do not expect to see the hexagonal phase in our magnetic field simulations if it is slightly favored energetically, because its symmetry group is incompatible with periodic boundary conditions in a rectangular system.  If it is highly favored energetically, we ought to see multiple hexagonal domains within the rectangular system.  This was not seen in our simulations.

\begin{figure}[t!]
\centering
\begin{minipage}{3.5 cm}
\includegraphics[width=1.0\textwidth]{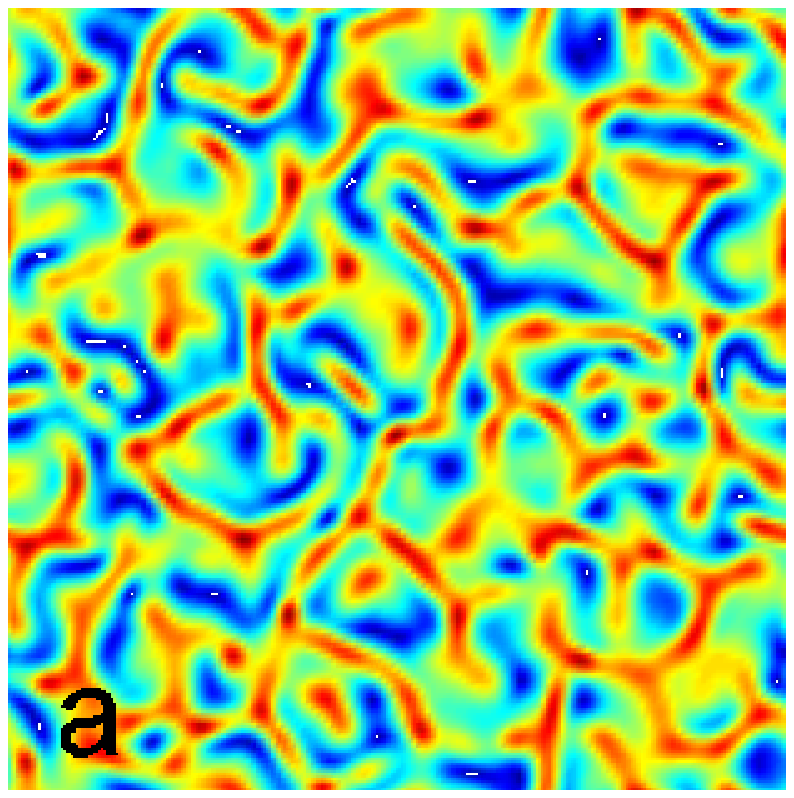}
\end{minipage}
\begin{minipage}{5 cm}
\includegraphics[width=1.1\textwidth]{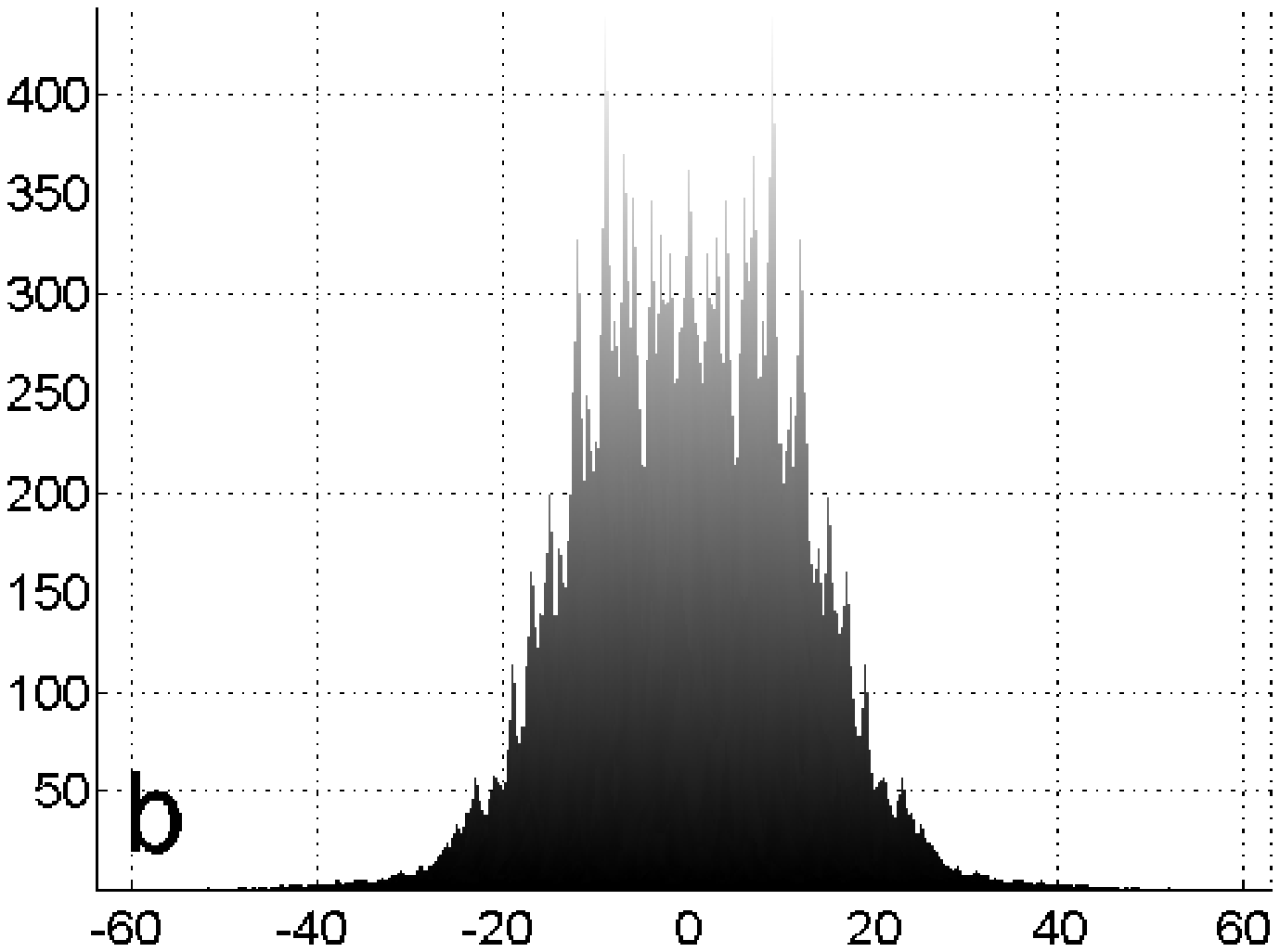}
\end{minipage}
\caption{(a) Ginzburg-Landau order parameter of the stripe solution in the absence of fields and (b) the amplitudes of Fourier coefficients.  There are no preferred orientations for the stripes in the isotropic CGLE, Eq. \ref{Equation2}.}
\label{NFStripe}
\end{figure}

Stripes formed in our simulations (shown in Fig. \ref{NFStripe}a) are mobile. Since we do not consider anisotropy our stripes are different from the highly parallel DOS modulations seen in STM experiments.  By introducing an anisotropic Laplacian or a pinning potential we are able to force the formation of parallel stripes which are then significantly less mobile than the isotropic ones. Because stripes seen in our simulations are not parallel we find the typical wavelength by analysing the Fourier spectrum. 
The spectrum of the field free case is shown in Fig. \ref{NFStripe}b. 

\begin{figure}[h!]
\centering
\begin{minipage}{3.5 cm}
\includegraphics[width=1.0\textwidth]{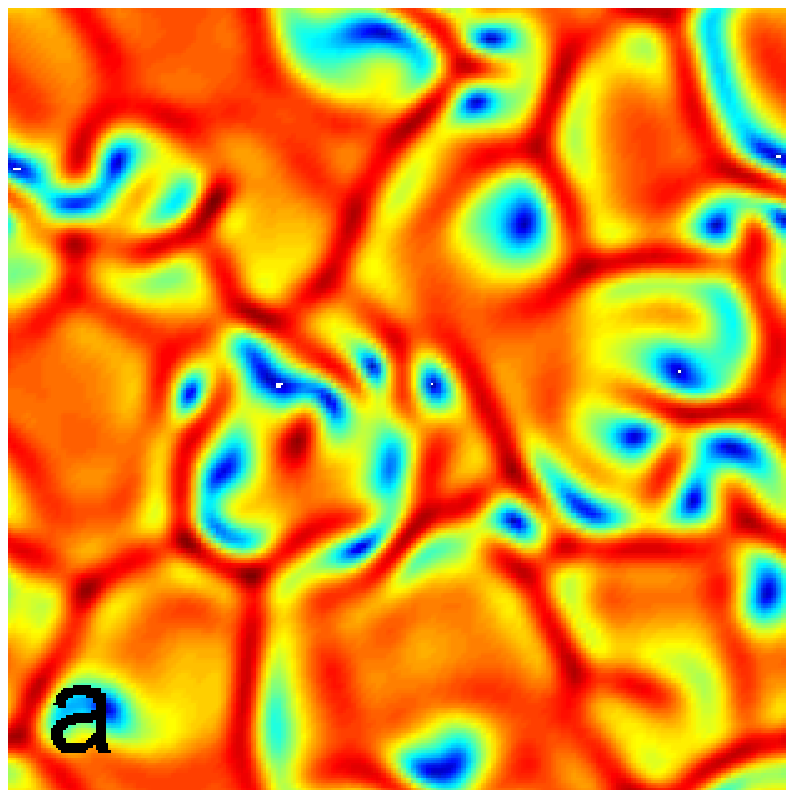}
\end{minipage}
\begin{minipage}{5 cm}
\includegraphics[width=1.1\textwidth]{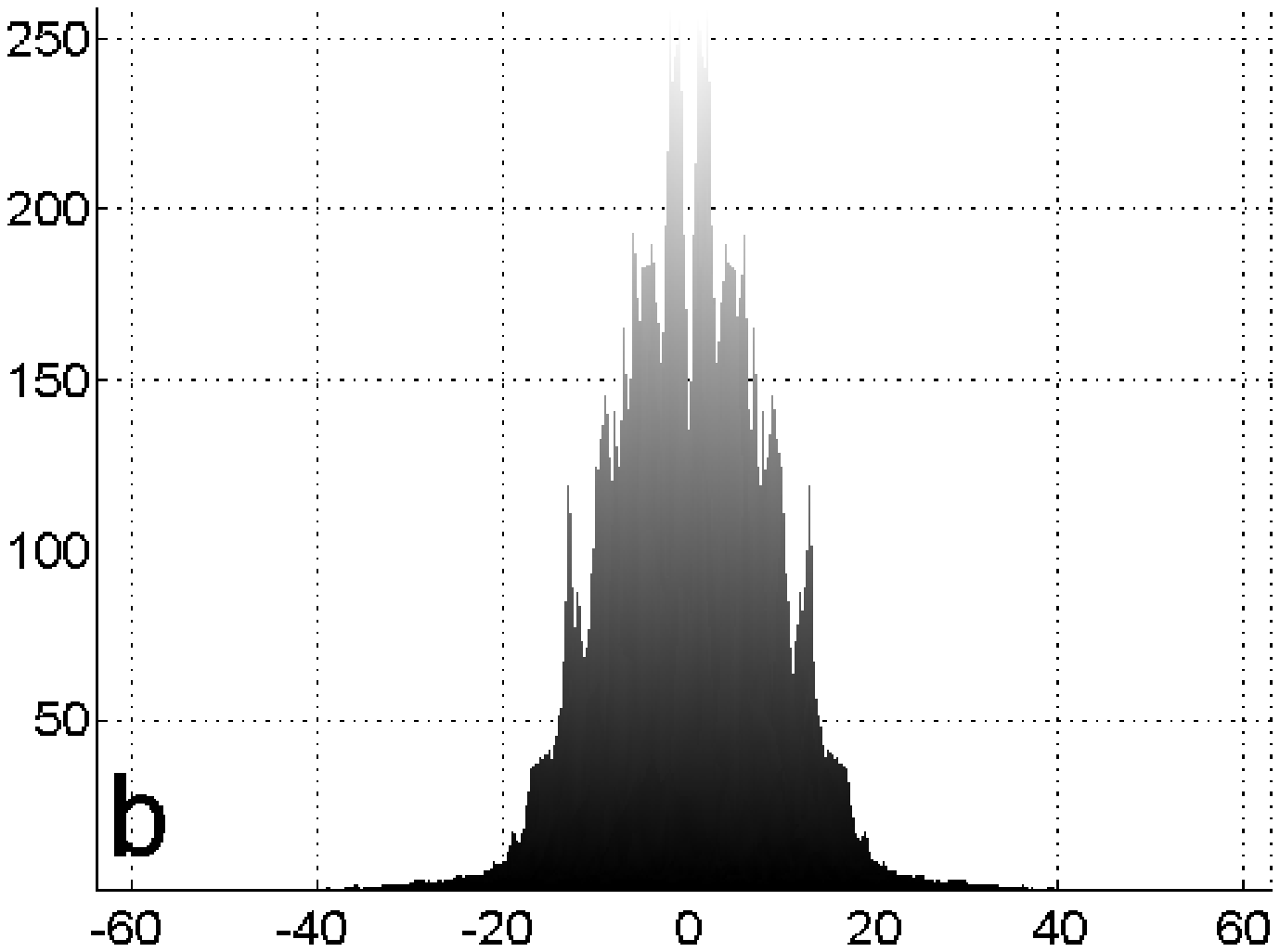}
\end{minipage}
\caption{(a) A hole-like Ginzburg-Landau order parameter under the influence of a negative bias voltage forms wider stripes. (b) The Fourier amplitudes then become shifted to smaller values compared to the field-free case.}
\label{WEFMIStripe}
\end{figure}

\begin{figure}[h!]
\begin{minipage}{3.5 cm}
\includegraphics[width=1.0\textwidth]{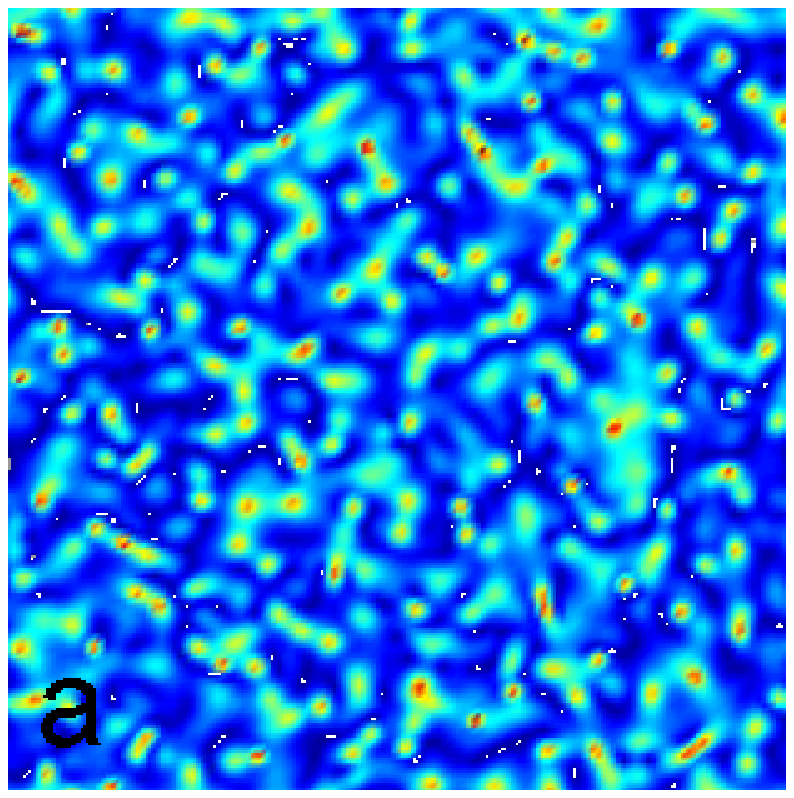}		
\end{minipage}
\begin{minipage}{5 cm}
\includegraphics[width=1.1\textwidth]{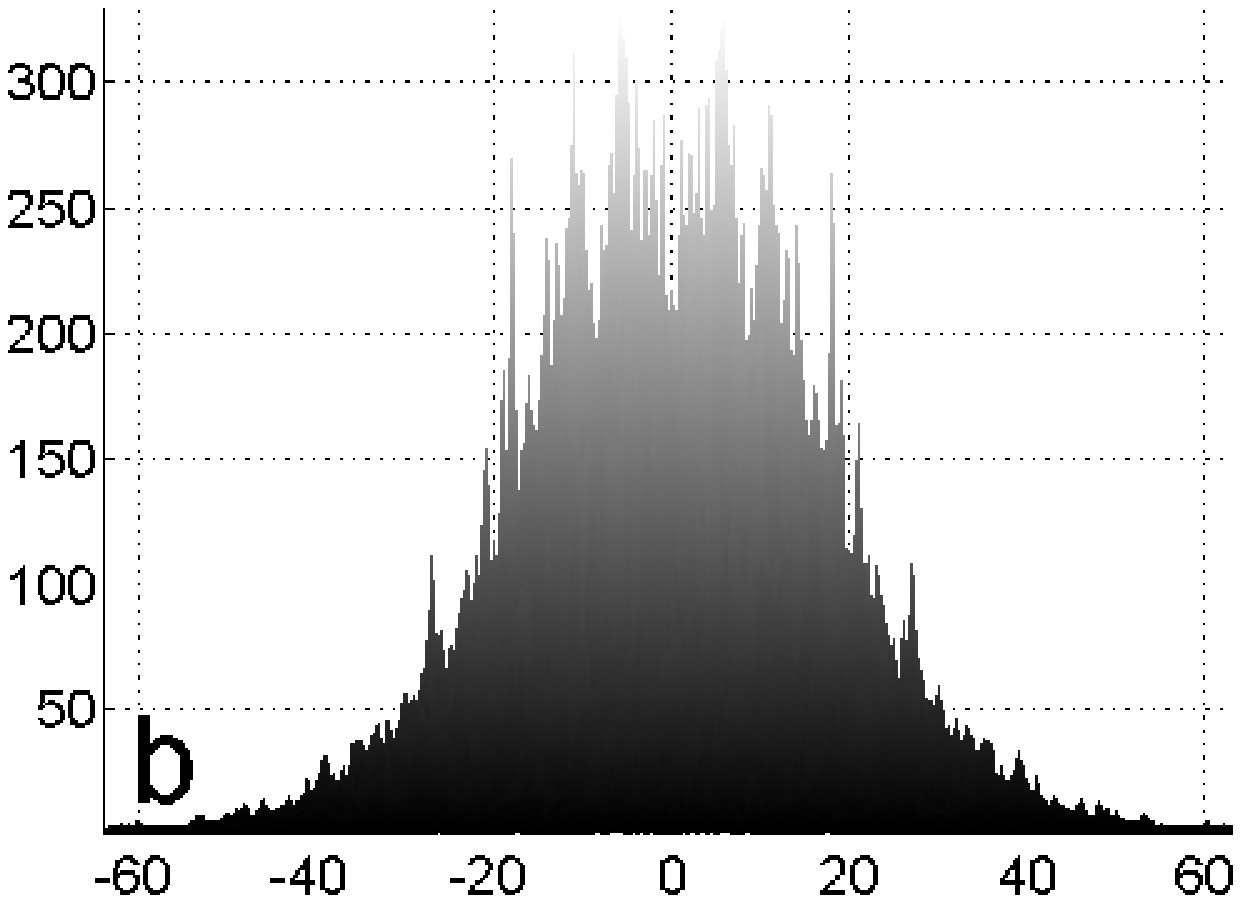}
\end{minipage}
\caption{(a) For an electron-like order parameter we found that stripes becomes narrower under the influence of a negative bias. (b) The amplitudes of the Fourier coefficients are then shifted to larger $k$-values.}
\label{WEFPLStripe}
\end{figure}

To investigate the influence of a STM tip on stripes we model the STM tip as a charged sphere with a radius of 20 nm at a distance 1 nm from the superconducting surface. In the presence of external electric field the initial nucleation phase is identical to the field-free case, if we start from the same random initial condition. The field strength only determines the critical size of the islands. For a positively(negatively) charged order parameter we found that islands reach smaller(larger) critical sizes under a negative bias voltage of a few pico-Volt. For a positive bias of the same strength the opposite is true. The width of stripes then depends on the critical size of the collapsing island. In general we found that larger (smaller) islands result in wider (more narrow) stripes (Fig. \ref{WEFMIStripe}a \& \ref{WEFPLStripe}a). The change of the Fourier coefficients for a hole- and an electron-like order parameter under a negative bias voltage can be seen in Fig. \ref{WEFMIStripe}b and \ref{WEFPLStripe}b respectively.

When we compare our simulations with results described by Hoffman et al. --- they see narrowing along ($\pi,\pi$) and widening along ($\pi,0$) --- we find that the dependence on the direction in $k$-space corresponds to electron- or hole-like character of our order parameter. A variety of experiments analysing the band structure of cuprates came to different conclusions. Whether the Fermi surface is electon-like and centred at ($0,0$) \cite{ELIKE} or hole-like with its centre at ($\pi,\pi$) \cite{HLIKE} is unclear. With our simulation we can qualitatively reproduce the change in the width of stripes as observed in Ref. \onlinecite{Hoffman} if we assume that the band structure is hole-like along the ($\pi,0$) direction and electron-like along the ($\pi,\pi$) direction.


\begin{figure}[h!]
\centering
\begin{minipage}{4.2 cm}
\includegraphics[width=1.0\textwidth]{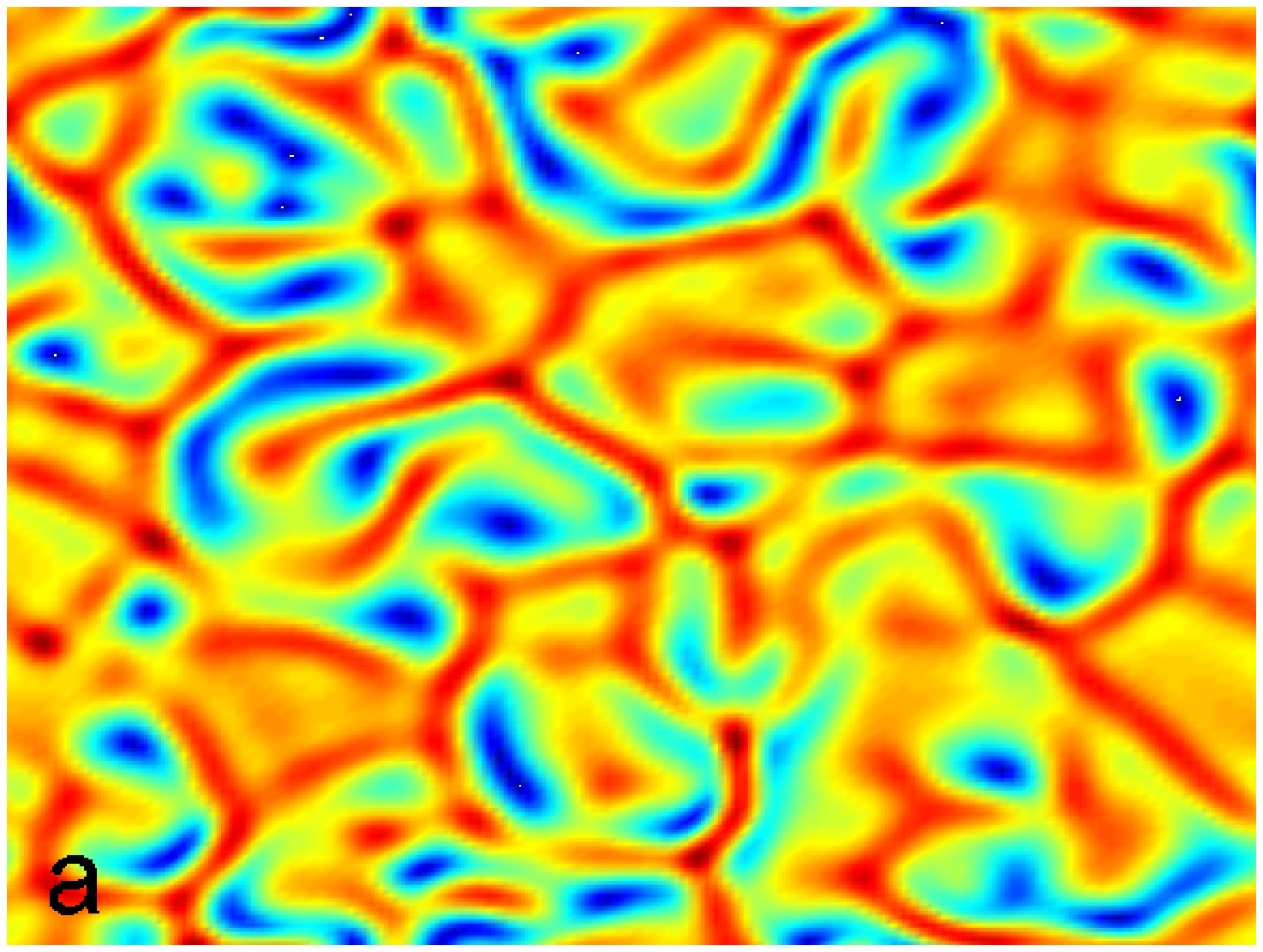}
\end{minipage}
\begin{minipage}{4.2 cm}
\includegraphics[width=1.0\textwidth]{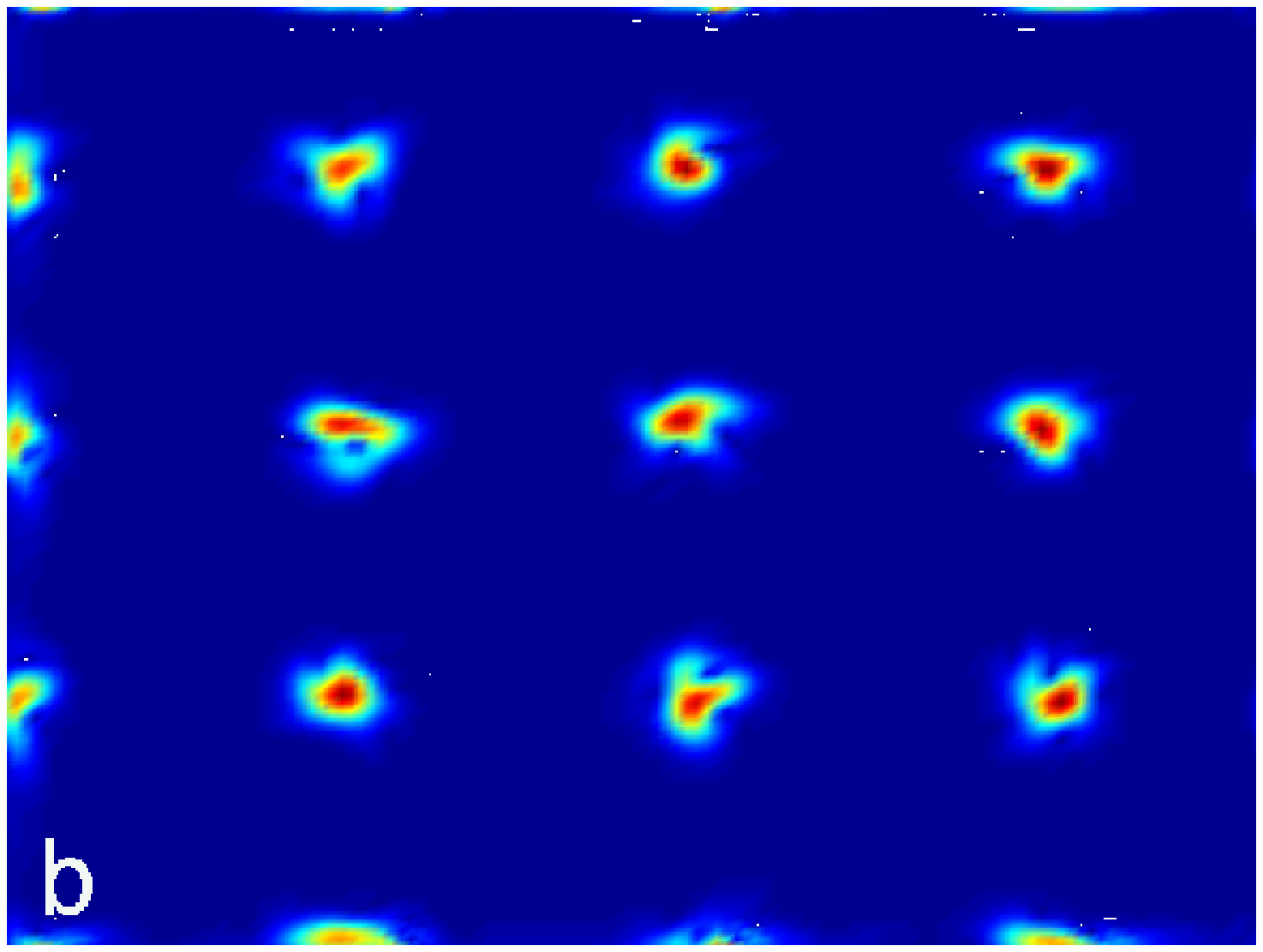}
\end{minipage}
\caption{Starting from well developed stripes, we see that a magnetic field destroys the stripe phase.  Within times that are short compared to the typical time scale of stripe dynamics, we observe the formation of uniformly-spaced vortex-like structures across the surface. Both figures show the absolute values of the numerical solutions.}
\label{MFStripe}
\end{figure}

Finally we check whether it is plausible to associate $\Psi$ with the superconducting order parameter in HTSC materials. Since strong magnetic fields destroy superconductivity, we ran simulations using the finite difference approach over a range of magnetic field strengths.
When we apply a perpendicular magnetic (of the order of pico-Tesla) starting from a random initial condition no stripes are formed and the solution of the CGLE becomes identically zero. We then turn on the magnetic field after stripes have formed (Fig. \ref{MFStripe}a), and observe that the stripes narrow until they break and form a uniform array of peaks (Fig. \ref{MFStripe}b). With time the peaks shrink until they become washed out completely. It is important to emphasize that these are not the vortices usually observed in type-II superconductors. We suspect that these features might have become the hexagonal phase if we would have used the correct boundary conditions.

To summarise, in this work we investigated the response of a charged order parameter with stripes to electric fields, by exploring numerical solutions of the CGLE.  For a hole-like order parameter under the influence of a negative bias voltage, stripes in our simulation widen.  The same trend is observed in STM measurment \cite{Hoffman} along the ($\pi,0$) direction.  With an electron-like order parameter we find narrower stripes --- a behaviour that is actually seen along the ($\pi,\pi$) direction in Ref. \onlinecite{Hoffman}.
We further solve the CGLE in the presence of a magnetic field, to show that both the homogeneous and stripe phases are destroyed, and thus it is plausible for $\Psi$ to be associated with a superconducting order parameter.
While we cannot claim that the CGLE accurately describes the superconducting order parameter in HTSC, we believe the qualitative response of the true order parameter to an external electric field will be similar.

This research is supported by startup grant SUG 19/07 from the Nanyang Technological University.  We acknowledge C. Panagopoulos and S. Uchida for useful discussions on the phenomenology surrounding HTSC.  We thank C. L. Henley for discussions on the stability of the stripe phase relative to the homogeneous superconducting phase, and also thank D. Wang for discussions on numerical methods for solving nonlinear PDEs.


\begin{thebibliography}{99}
\bibitem{footnote} During the nucleation stage of the simulations, the pseudo-spectral and finite difference solutions are nearly identical.  Only small variations arise because of the different nature of numerical errors in the two methods.  As the islands are breaking up into stripes, the finite difference method is not able to accurately cope with the fast changes in the order parameter $\Psi$.  However, after stripes have formed, and used as an initial condition, the two numerical methods are in good agreement with each other for a short period of time.
\end{thebibliography}
\end{document}